\documentclass[11pt,fleqn,twoside]{article}

\newif\ifreport
\reporttrue

\usepackage[german,english]{babel}
\usepackage{derireport}

\usepackage{times}
\usepackage{alltt}
\usepackage{epsfig}
\usepackage{url}
\usepackage{xspace}
\usepackage{latexsym}
\usepackage{rotating} 
\usepackage{pms}

\usepackage{graphicx}
\usepackage[tight]{subfigure}
\usepackage{verbatim}
\usepackage{epsfig}
\usepackage{amsfonts}
\usepackage{amssymb}
\usepackage{url}
\usepackage{cite}
\usepackage{colortbl}
\usepackage{tabularx}
\usepackage[T1]{fontenc}
\usepackage{ae,aecompl}
\usepackage{pslatex}
\usepackage[leftcaption]{sidecap}

\foreignreport

\title{Analysing Parallel and Passive Web Browsing Behavior and its Effects on Website Metrics}

\authornames{
Christian von der Weth
\and
Manfred Hauswirth
}

\innertitle{Analysing Parallel and Passive Web Browsing Behavior and its Effects on Website Metrics}

\author{
Christian von der Weth%
\affiliation{
DERI (Digital Enterprise Research Institute), National University of Ireland,
Galway\protect, IDA Business Park, Lower Dangan, Galway, Ireland.
\mbox{E-mail: christian.vonderweth@deri.org.}}
\and
Manfred Hauswirth%
\affiliation{
DERI (Digital Enterprise Research Institute), National University of Ireland,
Galway\protect, IDA Business Park, Lower Dangan, Galway, Ireland.
\mbox{E-mail: manfred.hauswirth@deri.org.}}
}

\published{
}

\acknowledgement{
This work has been funded (in a part) by
    Science Foundation Ireland (Grant no.~SFI/08/CE/I1380 -- L\'{i}on-2)
}

\abstract{
Getting deeper insights into the online browsing behavior of Web users has been a major research topic since the advent of the WWW. It provides useful information to optimize website design, Web browser design, search engines offerings, and online advertisement. We argue that new technologies and new services continue to have significant effects on the way how people browse the Web. For example, listening to music clips on \textsc{YouTube} or to a radio station on \textsc{Last.fm} does not require users to sit in front of their computer. Social media and networking sites like \textsc{Facebook} or micro-blogging sites like \textsc{Twitter} have attracted new types of users that previously were less inclined to go online. These changes in how people browse the Web feature new characteristics which are not well understood so far. In this paper, we provide novel and unique insights by presenting first results of DOBBS, our long-term effort to create a comprehensive and representative dataset capturing online 
user behavior. We firstly investigate the concepts of \textit{parallel browsing} and \textit{passive browsing}, showing that browsing the Web is no longer a dedicated task for many users. Based on these results, we then analyze their impact on the calculation of a user's dwell time -- i.e., the time the user spends on a webpage -- which has become an important metric to quantify the popularity of websites.
\\*[\parskip]
~
\\*[\parskip]
}

\deritr{2014-02-20}
\date{February 2014}

\begin{document}

%
% the next line MUST come first 
%

\maketitle

\newpage 
\pagenumbering{Roman}

{\small
%\vspace*{-3.5\baselineskip}

\tableofcontents
}

%\cleardoublepage

\newpage
\pagenumbering{arabic}

\section{Introduction} 
\label{sec:introduction}
The investigation of the browsing behavior of Web users provides useful information to optimize website design, Web browser design, search engines offerings, and online advertisement. This has been a topic of active research since the Web started and a large body of work in this area exists. We argue, however, that new online services as well as advances in Web technologies have changed and continue to change the meaning behind ``browsing the Web''. This requires a fresh look at the problem, including the question whether the used models are still appropriate.

Today's bandwidth resources and the resulting success of new kinds of online platforms changed the usage of the Web significantly. Particularly media or streaming sites like \textsc{YouTube}, \textsc{Netflix} or \textsc{Last.fm} allow users to watch video clips or movies, or listen to online radio. Given these trends, browsing the Web has become more and more a passive activity where users visit a website but do not necessarily interact with that site (all the time). These new types of online platforms and services also have a significant effect on the demographics of the Web. Social network and social media sites like \textsc{Facebook}, micro-blogging sites like \textsc{Twitter}, the omnipresence of online shops, online browser games, etc. attract new groups of users, including less tech-savvy ones that previously were less inclined to frequently use the Web, if at all. Furthermore, as studies show (see Section~\ref{sec:relatedwork}), particularly social network sites have a strong sociological impact, 
indicated by the increasing time users spend online and how they arrange their social life accordingly. 

Besides new types of services, also new technologies have changed how we use the Web. Technologies such as Ajax (Asynchronous JavaScript and XML)~\cite{Holdener08Ajax} and WebSockets~\cite{Fette11WebSockets} provide methods for updating content on a page without reloading the whole page (or going to a new page) by requesting only small bits of information from a Web server instead of entire content. The involved benefits of such (incremental) updates -- particularly the reduced bandwidth consumption and making Web applications behave more like desktop applications -- spurred the adoption of such technologies by many popular websites and online platforms. In the last two decades, Web browsers have developed from simple tools to render HTML source code to powerful and sophisticated application platforms. While a lot has been done ``under the hood'' -- for example, the support of all kinds of Web standards, media formats, etc., or performance optimization techniques like caching of webpage content -- browsers 
also have improved with respect to their usability. For example, concepts such as tabbed browsing or bookmarking are featured in all modern browsers.

Such changes in how people browse the Web feature new characteristics which are not well understood so far. This is largely due to the lack of meaningful datasets capturing user behavior. Server-side data sources such as Web server access logs or search engines transaction logs do not provide sufficient information (e.g., the time a user stayed on a page or the usage of multiple tabs or browser windows). On the other hand, client-side studies are typically conducted as lab experiments. While this approach is valid to investigate very specific user behavior, it is very unlikely to elicit the every-day browsing behavior of users. To address these limitations, we developed DOBBS (DERI Online Browsing Behavior Study)~\cite{vdw13DOBBS}. DOBBS is a long-term effort to create such a comprehensive and representative dataset in a non-intrusive, completely anonymous and privacy-preserving way. DOBBS provides add-ons for Firefox and Chrome that log all major events in the context of browsing the Web, such as the 
opening 
of new tabs, the loading of new pages, and many more. The data about each event is sent to a central repository, and the resulting dataset is made public to be downloaded on a regular basis. The goal is to allow researchers and software engineers to perform better analyses and to gain deeper insights into the browsing behavior of Web users.

In this paper, we present a first evaluation of the DOBBS dataset. Since browsing behavior may refer to a large variety of aspects, we focus on those we deem not sufficiently addressed by existing works, if at all. Firstly, we investigate users' level of parallel browsing, i.e., their usage of tabbed browsing or multiple browser windows at the same time. Our results show that most users open only one browser window but use multiple tabs basically all the time. Secondly, we analyze the phenomenon of passive browsing, i.e., the time of inactivity during a user's browsing session. For this, we use different means to calculate the idle time of users and measuring their effects on the results. In general, the longer a browsing session the longer the time user was idling. This indicates that browsing the Web is no longer a dedicated task. And thirdly, we take an in-depth look into users dwell time on individual pages, considering whether the page was actually 
visible (e.g., the page was not in a background tab) and the user was indeed actively browsing. We show that incorporating these aspects has a significant effect on the calculation of users' dwell time. This in turn challenges current approaches using dwell time as metric to quantify the popularity, usefulness or relevance of websites.

Paper outline:
Section~\ref{sec:relatedwork} reviews related to put our contributions into context.
Section~\ref{sec:dobbs} gives a brief overview to DOBBS, focusing on important design decisions and technical details that affect the quality of the resulting dataset.
Section~\ref{sec:evaluation} describes our applied data cleaning steps and presents the results of our evaluation.
Section~\ref{sec:discussion} provides a critical discussion of the results and motivates their practical relevance.
Section~\ref{sec:conclusions} concludes this paper.

\section{Related Work}
\label{sec:relatedwork}
The investigation of browsing behavior has been major research field since the early 90's. As many studies show, with the advent of Web 2.0, the way people browse the Web has changed significantly. Many of the now most popular websites went online in the middle of the first decade of 2000 (e.g., \textsc{Facebook} (2004), \textsc{LinkedIn} (2003), \textsc{YouTube} (2005), \textsc{Reddit} (2005), or \textsc{Twitter} (2006)). We therefore limit ourselves to rather recent studies, i.e., not before 2005. We categorize related work with respect to the way the study has been conducted, i.e., how the data has been collected.
\\
\\
\textit{Server-side studies.} One type of data source are Web server access logs~\cite{Grace10WebLogData}. Their usefulness is typically limited to specific research questions since they generally only report on user actions within a single site. The results in \cite{Xue10UserNavigation} show that users often exhibit different behavior patterns rather than a single one when browsing for information. In \cite{Hawwash10MiningAndTracking} the aim was to investigate how the browsing behavior of users changes over time. \cite{Meiss09WhatsInASession} collected the server accesses on a university router and thus fetching/intercepting the requests to all sites from the intranet to the Web. The results confirm the long-tailed distributions in site traffic. A second type of server-side data sources are search engine transaction logs. Their analysis provides insights into query behavior and the selected elements of the result list. However, the navigation paths of users after leaving the result page(s) remain unknown. \
cite{Agichtein06ImprovingWebSearch} investigated how user behavior, derived from click streams recorded within the \textsc{MSN} search engine, can be used as implicit feedback to improve the ranking of query results. The authors of \cite{Beitzel07TemporalAnalysis} analyzed an \textsc{AOL} query log in terms of how the query behavior of users changes over time. They found that certain topical categories can exhibit both short-term and long-term query trends. Using query logs of the \textsc{Yahoo!} search engine, \cite{Wedig06LargeScaleAnalysis} found that after a few hundred queries a user's topical interest distribution converges and becomes distinct from the overall population. 
\\
\\
\textit{Client-side studies.}
Collecting data on the client side, in general, requires users to install browser extensions (or use special browsers) that log all user actions. This approach enables to capture browsing behavior in much more detail. \cite{Goel12WhoDoesWhat} focused on demographic factors, i.e., how age, sex, race, education, and income affectwhich sites and for how long users are visiting. \cite{Kellar06TheImpact} investigated how the browsing behavior of users depends on the task (e.g., fact finding, information gathering, etc.). In \cite{Adar08LargeScaleAnalysis} the authors identify twelve different types of revisitation behavior, based on which they outline recommendations towards the design of Web browsers, search engines, and websites. \cite{Weinreich08NotQuite} investigated the use of parallel browser windows or tabs to navigate between pages, showing that different users show very characteristic behaviors. \cite{Kumar10Kumar} provides a taxonomy of page views: (a) content ($\sim 50\%$), like news, portals, 
games, verticals, multimedia, (b) communication ($\sim 33\%$) like email, social networking, forums, blogs, chat, and (c) search ($\sim 17\%$) like as Web search, item search, multimedia search. \cite{Dubroy10AStudyOfTabbed,Zhang11MeasuringWebPage,Huang12NoSearchResult} investigated tabbed browsing, i.e., the effect of multiple tabs within a browser window. Their results show that tabbed browsing is very popular and is speeding up the browsing process. 
\\
\\
\textit{Exploiting users' dwell time.} Various works investigated the time users' spent on webpages as implicit feedback or metric to quantify the popularity of pages or websites. \cite{Agichtein06ImprovingWebSearch,Bilenko08MiningTheSearchTrails,Chapelle09DynamicBaysianNetwork} aim to exploit dwell time as an additional parameter to generate a ranking of search results. \cite{Kelly04DisplayTime,White06StudyOnTheEffects} studied the correlation between the dwell time and the usefulness of documents in information-seeking tasks. The \textsc{BrowseRank} algorithm~\cite{Liu08BrowseRank} incorporates user behavior to rank the relevance of webpages and to combat spam. Works like~\cite{Nunez-Valdez12ImplicitFeedback,Yin13SilenceIsEverything} consider dwell time as implicit rating/voting to improve recommender systems. Here, the rating of a product derives from the time users spent on a product webpage. In a similar line, \cite{Hu08CollaborativeFiltering,Xin11MultiValue} try to use information about what people 
were watching and for how long to recommend TV shows or programs. Various works aim to model users' dwell time. While~\cite{Liu10UnderstandingWebBrowsing} use a Weibull distribution, \cite{Yin13SilenceIsEverything} found that a log-Gaussian distribution provides better fits to model dwell times.
\\
\\
Summing up, server-side collected data fall short to sufficiently capture users' browsing behavior. Controlled/supervised studies conducted in a lab under time constraints are limited to investigate user behavior while solving a specific task. Lab studies always bring ``ordinary people in extraordinary situations'' which does not elicit normal behavior. Closest to our approach is the Web History Repository Project~\cite{Herder11WHR} which also features a browser add-on to capture browsing events. Efforts to use dwell as a metric rely on datasets that mostly provide only estimates regarding the times users spent on webpages. In contrast, we argue that DOBBS, as long-term field study, provides a uniquely detailed view into users browsing behavior. In~\cite{vdw13DOBBS} we first introduced DOBBS, and presented some exemplary analysis results on a very small sample to highlight the potential benefits of the resulting dataset. In this paper, we focus on the notions of parallel browsing and passive browsing. We 
show how these phenomenons affect the calculation of users' dwell time, which in turn affects the application of dwell time as metric to quantify the popularity of websites.

\section{The DOBBS System}
\label{sec:dobbs}
The DOBBS system uses a browser add-on\footnote{See project website: http://dobbs.deri.ie.} that captures browsing events and sends them to a central server. Browsing events comprise, for example, the adding/closing of tabs, the loading of webpages, window status changes, and user activity changes. The DOBBS add-on is an ``install-and-forget'' application, running silently in the background. The rationale is to avoid affecting users' browsing behavior. The add-on implements a \textit{best-effort logging} -- that is, it tries to sends each recorded event immediately to the server. In case of connection errors, the add-on simply ignores this unsuccessful attempt, discards the data, and continues trying to send subsequent events as if nothing happened. With this approach, we addressed the trade-off between handling all exceptions and the degree of complexity of the add-on in favor of the latter. In the very most cases, this is acceptable since users are unlikely to continue browsing if they lose their Internet 
connection. Still, as a result, the dataset is inherently imperfect, typically in form of missing events but also, in rather rare cases, duplicate events. We discuss the data cleaning techniques we applied prior   to our evaluation in Section~\ref{sec:evaluation}.

\subsection{Recorded Events}
\label{sec:data}
The main unit of information within DOBBS is an event. The dataset distinguishes between \textit{window events}, \textit{session events}, and \textit{browsing events}, which we describe in the following. Beside event-specific attributes, all event types share a set of common attributes (see Table~\ref{tab:core_attributes}). The complete list of all recorded events and their representations within the dataset can be found on the project website.
\begin{table}
\begin{center}
\begin{tabular}{|l|p{9cm}|}
  \hline
  \textbf{Attribute} & \textbf{Description} \\
  \hline
  \texttt{time} &  time on client side when an event has occurred\\
  \hline
  \texttt{tz\_offset} &  difference between UTC time and client time, in minutes (e.g., for GMT+2, \texttt{tz\_offset} = -120)\\
  \hline
  \texttt{user\_id} &  unique numeric identifier of a user, randomly generated during the installation of the add-on\\
  \hline
  \texttt{window\_id} &  unique numeric identifier of a browser window, randomly generated at the time of opening\\
  \hline
  \texttt{session\_id} &  unique numeric identifier of a logging session, randomly generated at session start\\
  \hline
  \texttt{tab\_id} &  numeric identifier of an open browser tab, unique within each browser window\\
  \hline
\end{tabular} 
\end{center}
\caption{Core attributes that are logged for all events} 
\label{tab:core_attributes}
\end{table}
\\
\\
\textit{Window events.}
Window events encompass all events that are associated with interacting with a browser window, e.g., the opening and closing of a browser window or individual tabs within a window. Both windows and tabs feature a unique identifier. The add-on captures any change in the state of a window (maximized, minimized, normal, full screen). The add-on also keeps track if a browser window lost focus, i.e., became a background window on the user's desktop, or regained the focus again. 
\\
\\
\textit{Session events.}
A session denotes the interval in which all occurring events are recorded and sent to the server. Users can end a session explicitly or implicitly by entering the Private Browsing / Incognito mode. Analogously, switching the logging process back on or leaving the Private Browsing / Incognito mode initiates a new session. Each session features a unique identifier. We also consider the activity state of a user (\textit{active} or \textit{inactive}) as a session event. Both Firefox and Chrome implement a feature that check the activity of a user every few seconds and fire an event if that state changes. As a design decision, a user is considered as inactive if the user was not active for at least one minute. 
\begin{table}[t!]
\begin{center}
\begin{tabular}{|c|p{8cm}|}
  \hline
  \textbf{Level} & \textbf{Example} \\
  \hline\hline
  domain   & \texttt{example.org} \\
  \hline
  (sub-)domain   &\texttt{topic.example.org} \\
  \hline
  full path   &\texttt{topic.example.org/dir/index.php} \\
  \hline
  full URL   &\texttt{topic.example.org/dir/index.php?id=42} \\
  \hline
\end{tabular} 
\end{center}
\caption{Considered components of a URL} 
\label{tab:url_components}
\end{table}
\\
\\
\textit{Browsing events.}
Browsing events are associated with navigating between webpages. This includes new page loads in the selected or a background tab, but also the state of visibility of a webpage. A page becomes visible after it has been loaded in the selected tab, or the background tab containing the page becomes the selected tab. Analogously, a page becomes invisible before it is unloaded in the selected tab, or the tab containing the page is moved in background. The add-on also captures the cause of the events (e.g., click on link or bookmark). Browsing events carry the information about the visited pages, i.e., their URL, as sensitive information. To address the trade-off between privacy preservation and the possible insights into browsing behavior, the add-on distinguishes four different ``levels'' for each URL: the domain, the domain and all subdomains, the full path, and the full URL itself. Table~\ref{tab:url_components} shows an example. The add-on encrypts each component individually before sending the 
complete record to the server. This allows to group browsing events according their shared domain, subdomain, etc., and not only according to the complete URL.

\subsection{Privacy Preservation}
\label{sec:dobbs-privacy}
Naturally, logging user behavior in such a detailed fashion raises privacy concerns, potentially discouraging users to contribute. DOBBS therefore applies a whole set of techniques to ensure users' privacy as much as possible.
Firstly, the add-on anonymizes the logged data, with participants being identified only by a randomly generated integer value, and without any connection to their real-world identities. No information that may point to the real-world identities of participants such as IP addresses or explicitly requested email addresses are ever collected or transferred to the server.
Secondly, all sensitive data -- that is, the URLs (and its components; see Table~\ref{tab:url_components}) of the webpages the participants were browsing on -- are first encrypted on the user side and then sent to the server. 
Thirdly, via an entry in the browser's menu, a participant can manually stop and resume the logging process at any time. The add-on also respects Firefox's Private Browsing as well as Google Chrome's Incognito mode, i.e., the logging is suspended during these modes.
And lastly, apart from the events as described above, no other information are logged. This includes manual user input for, e.g., the optional search field in the toolbar, but particularly for any kind of form fields embedded in webpages, e.g., for user names or passwords.

\section{Evaluation}
\label{sec:evaluation}
This section presents the results of our analysis of the DOBBS dataset as it was available in January 2014. We first describe our steps to clean the dataset, define the scope of our evaluation, and describe our basic methodology.
\\
\\
\textit{Data cleaning.}
Given that DOBBS is an unsupervised field study and the add-on implements a best-effort logging, the dataset contains entries the potentially affect an analysis. We therefore performed various data cleaning steps as countermeasures. Firstly, a lot of users just tried the add-on but did not continued to use it. For our evaluation, we identified the 30 most active users with an activity ranging from 1,6k up to 88k overall page loads for individual users. We then removed all data not stemming from this set of active users. We also remove corrupt data in terms of duplicates. For example, some sessions feature two SESSION CLOSED events. In such cases, we keep only the first occurrence of an event. We want to point out, however, that the number of duplicates is very low.

In a second step, we tried to handle missing data. The causes for the loss of recorded events range from connection problems, browser crashes to specific user behavior. Regarding the latter, as soon as a browser is not normally closed -- e.g., by forcefully killing the corresponding process or by shutting down the computer -- the add-on cannot send SESSION CLOSED, WINDOW CLOSED and related events. While our observations show that is only the case for a rather small number of sessions on the whole, it potentially excludes the data from participants that mostly or always shutdown the computer without closing their browser. We therefore estimated the times a window and session have been closed as time of the last recorded event associated to a session (e.g., time of the last page load). Finally, we added these estimates into the dataset.
\\
\\
\textit{Scope of evaluation.}
The DOBBS datasets allows the analysis of users' browsing behavior from a wide range of perspectives and different levels of detail. In this paper, we focused on the following three aspects we deem not only of interest but also not addressed in a similar fashion by any existing work:

\textit{(1) Parallel browsing:} How do user make use of multiple browser windows or of the now very popular feature of tabbed browsing to open and view multiple web pages in parallel?

\textit{(2) Passive browsing:} How common is the phenomenon of passive browsing, i.e., the phases in which users do not actively interact with their open browser windows and loaded pages?

\textit{(3) Website popularity:} How does the ranking of websites regarding their popularity change with the effects of parallel and passive browsing activity taken into account?
\\
\\
\textit{Evaluation methodology.} DOBBS has first been announced and promoted in August 2014 over popular mailing lists. The study is a long-term effort, with new users continuously joining over time. As a result, the time users have contributed to the dataset differ significantly. It is therefore not meaningful to compare absolute values such as the number or accumulated durations of browsing session. We present the results of our analysis in form of averages, relative values or correlations that are independent from the time users have joined DOBBS. We observed that the distribution of most of the parameters, e.g., the length of sessions, is skewed, even for individual user. Thus, if not stated otherwise, we always use the median to calculate average values.

\begin{figure}[t!]
\centering
\subfigure[Comparison between the usage of tabbed browsing and multiple browser windows (sorted according to average number of open tabs).]{
\includegraphics[width=0.7\textwidth]{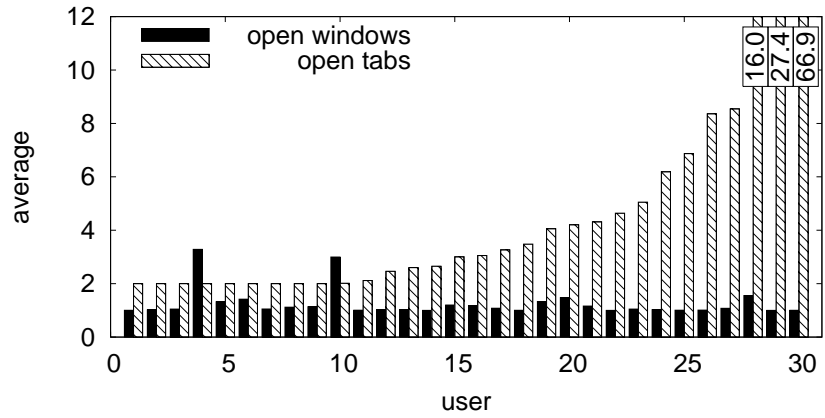}
\label{fig:average-open-windows-tabs}
}
\subfigure[Distribution of the number of open browser windows.]{
\includegraphics[width=0.7\textwidth]{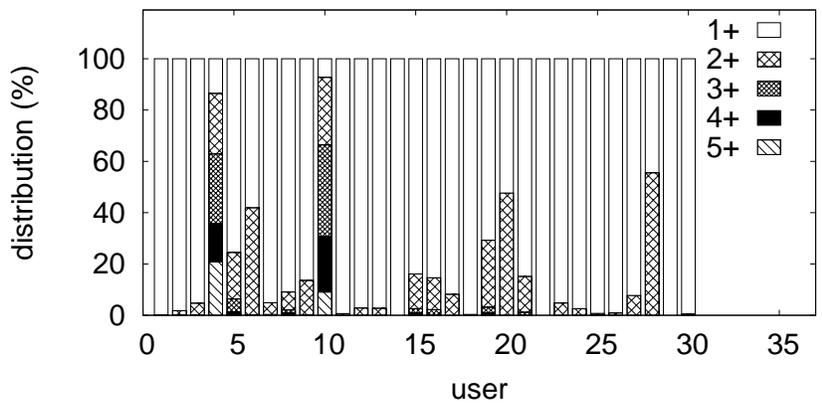}
\label{fig:window-overlap}
}
\subfigure[Distribution of the number of open browser tabs.]{
\includegraphics[width=0.7\textwidth]{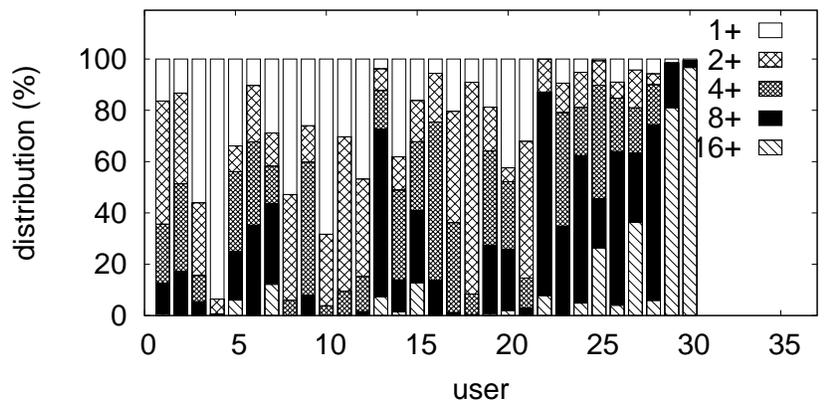}
\label{fig:tab-overlap}
}
\label{fig:parallel-browsing}
\caption[]{Result of analyzing the parallel browsing behavior.}
\end{figure}

\subsection{Parallel browsing}
All modern browser allow to open multiple windows in parallel and support tabbed browsing, i.e., opening multiple tabs within one browser window. From a user's perspective, the fundamental difference is that multiple browser windows facilitate viewing of different web pages at the same time. With tabbed browsing only a single page at a time is visible, and users have to switch between tabs to view the different loaded pages. Naturally, multiple open windows and tabbed browsing can be used in combination.

Figure~\ref{fig:average-open-windows-tabs} shows the average number of parallel open windows and parallel open tabs for all 30 users, sorted according to the value for open tabs. We calculated the average number of parallel open windows based the time intervals a user had one, two, three, etc. windows were open at the same time. Regarding the average number of parallel open tabs we first calculated this number for each individual session based on the time intervals one, two, three, etc. tabs were open at the same time. We then used the median the get the average for each user. The results show that tabbed browsing is way more common than open multiple browser windows. Except for two users, the average numbers of parallel open windows is less than 2. Tabbed browsing, on the other hand, turns out to be a commonly used feature with some users having a very high average number of open tabs.

Figures~\ref{fig:window-overlap} and~\ref{fig:tab-overlap} show the usage of multiple browser windows and tabbed browsing in more detail. In both figures, the position of each users in the graphs are in line with Figure~\ref{fig:average-open-windows-tabs}. To give an example of how to read the figures: User 5 is typically browsing with one browser window. $\sim$25\% of the time s/he is using at least two windows in parallel, and three or more windows $\sim$5\% of the time. Regarding tabbed browsing, $\sim$70\% of the time User 5 has at least two parallel open tabs, $\sim$60\% of the time at least four parallel open tabs, $\sim$30\% of the time at least eight parallel open tabs, and $\sim$5\% of the time at least 16 parallel open tabs. While we are not surprised that tabbed browsing is very common, we did not expect that having 8 tabs or more open at the same time has such a high likelihood.

\begin{figure}[t!]
 \centering
 \includegraphics[width=0.8\textwidth]{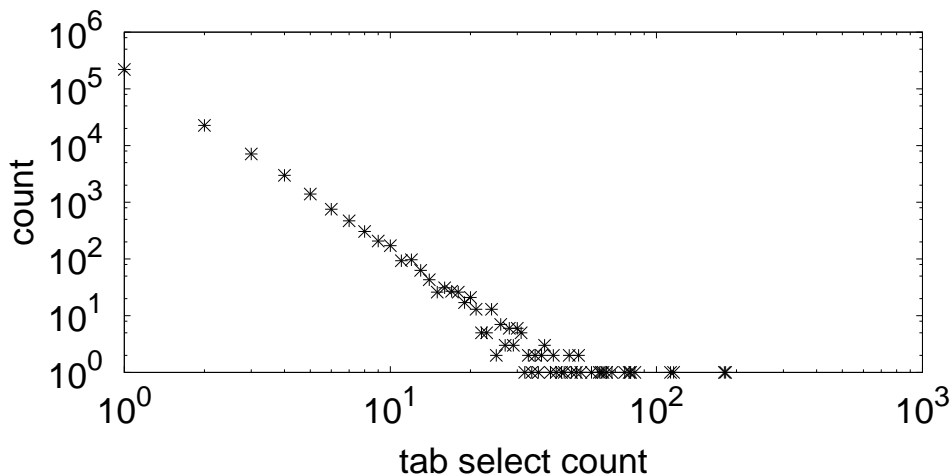}
 \caption{Distribution of the number a tab with a loaded page has been selected (after being a background tab).}
 \label{fig:dobbs-tab-select-distribution-all}
\end{figure}

\begin{figure}[t!]
 \centering
 \includegraphics[width=0.8\textwidth]{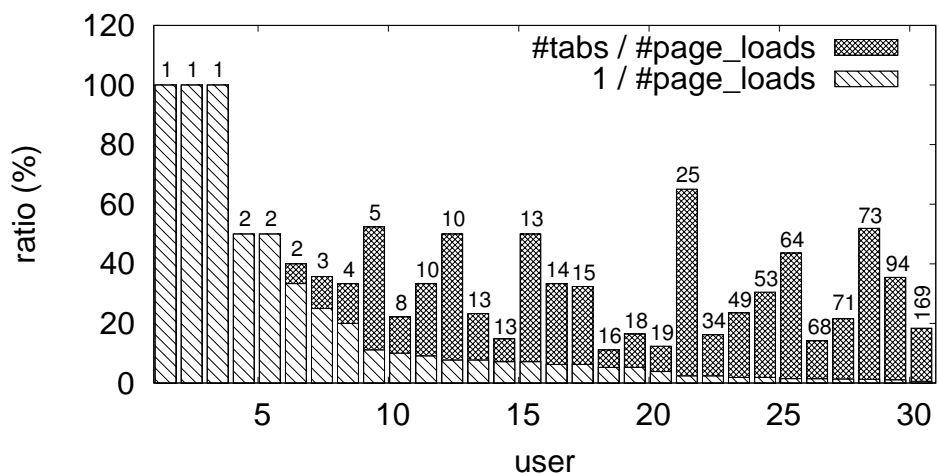}
 \caption{Ratio between average number of used tabs (median) and average number of page loads (median) per session.}
 \label{fig:dobbs-tab-page-load-ratio}
\end{figure}

Overall, the results show that parallel browsing is very common, with tabbed browsing the by far most popular way to do so. However, the degree of parallel browsing can vary significantly, particularly regarding using multiple tabs at the same time. It is interesting to see that users do not seem to mix multiple browser windows and tabbed browsers. Users 4 and 10, the only two users who regularly have more than one window open at the same time, are far less likely to open multiple tabs. As some kind of extreme cases, some users have a very high average number of open tabs, up to more than 60; cf. Figure~\ref{fig:average-open-windows-tabs}. Even more, Users 29 and 30 have more than 16 tabs open for most of their browsing time. We presume (a) that these users use the common feature of modern browsers to restore the previous browsing session at start-up, and (b) that these users might have installed available browser add-ons that allow for a more user-friendly tabbed browsing, e.g., by organizing subset of tabs 
into meaningful groups.

Beside the number of tabs open at the same time, parallel browsing also refers to the behavior of switching between open tabs. Figure~\ref{fig:dobbs-tab-select-distribution-all} shows the distribution of how often users selected a tab containing a specific webpage. Not surprisingly, the distribution shows a power law relationship. Most of the time, users open and select a new tab, use it for a while, and then close it again. Still, users often go back to loaded pages by switching to the corresponding tabs. Another interesting result -- no shown in the figure due to the log scale -- is that more than 6\% of pages that have been loaded was never visible. The reason for this is that users occasionally load a new page into a background tab but close this tab before they actually had a look at the page. Most prominently, when using a Web search engine, users often open the, e.g., top 5, result pages in background tabs in one go. Then, one by one, they switch between tabs to check each result. If one of the first 
pages they view provides sufficient information, they often simply close the tabs containing the remaining result pages.

In a last test, we investigated if and to what extent users ``re-use'' open tabs to load webpages. For this, we calculate the ratio between the average number of used tabs and average number of page loads per session. For both values we used the median to calculate the average. Figure~\ref{fig:dobbs-tab-page-load-ratio} shows the results, with the users sorted according to the average number of page loads per session (number above each bar). The lower the ratio the more do users tend to use one tab for multiple page loads. The height of the striped bars represents the minimal ratio $1/k$ (with $k$ being the number of page loads) in case a user would have used only the initial tab to load all pages. In general, the longer a browsing session in terms of page loads, the more do users tend to re-use tabs to load multiple pages. However, the results also show that this behavior differs noticeable between different users. For example, while User 14 and 15 open the same number of pages per session (here: 13), on 
average, User 15 opens them in new tab $~\sim$50\% of the time. User 14, on the other hand, uses far less tabs (but still more than one).

\subsection{Passive Browsing}
As outlined in the introduction, we argue that online browsing is for many users no longer a dedicated task. For example, users can watch a video clip or listening to online radio while writing a document or are busy with something completely different (cleaning, cooking, etc.).

\textit{Explicit idle time.} Both Firefox and Chrome check every few seconds whether a user is actively using the browser or not, and fire respective events if the activity state of a user changes. The explicit idle time is the difference between the pair of events representing the time a user entered and left the state of inactivity. The state is measured globally, i.e., a user is only considered idling if the user is inactive in all parallel open browser windows. As a design decision for DOBBS, a user is considered idling only after at least 1 minute of inactivity.

\textit{Implicit idle time.} Since each logged event has a timestamp, the data also allow us identifying phases of inactivity implicitly by the prolonged absence of any new event. Throughout our evaluation, we used three threshold values (60s, 240s, 960s) as minimum time that has to pass before a user is considered idling. We calculate the implicit idle time for each individual browser window, even if a user has multiple windows open in parallel.
\begin{figure}
 \centering
 \includegraphics[width=0.8\textwidth]{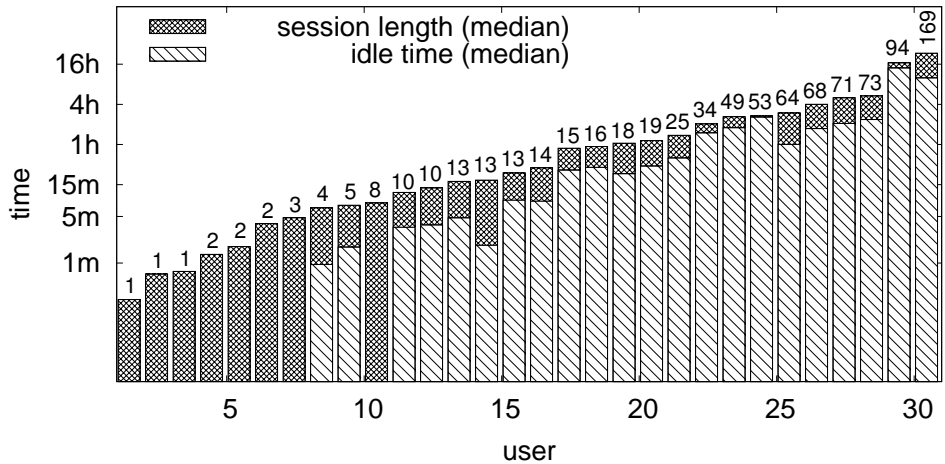}
 \caption{Average session length, average idle time and average number of page loads.}
 \label{fig:dobbs-session-length-and-idle-time}
\end{figure}
\begin{figure}
 \centering
 \includegraphics[width=0.8\textwidth]{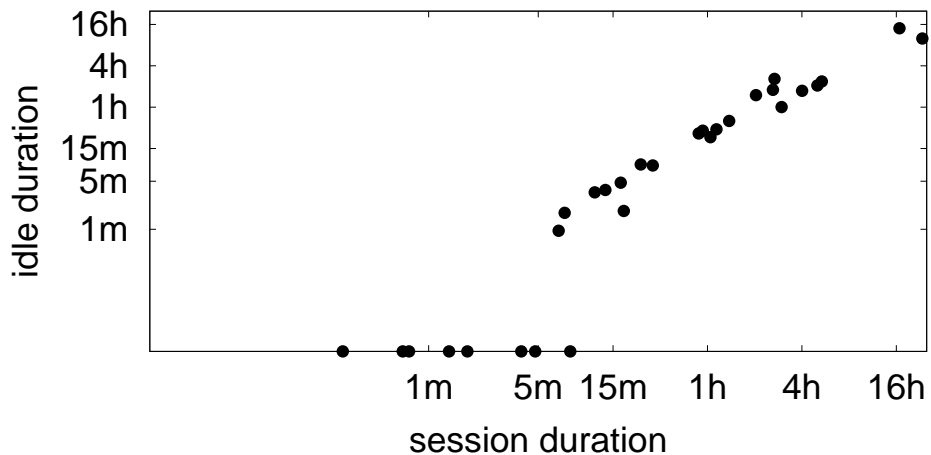}
 \caption{Correlation between average session length and explicit idle time for all users.}
 \label{fig:dobbs-median-correlation-full-vs-idle-time-explicit}
\end{figure}
\\
\\
Figure~\ref{fig:dobbs-session-length-and-idle-time} shows the average session length and the average explicit idle time for each user, as well as the average number of page loads as number over each bar. All three results are the median of the respective measures and are sorted according to the average session length. Several points are noteworthy. Firstly, even for this rather small set of users, the average session lengths vary significantly, ranging to sessions of up to several days. However, with increasing session length the idle time increases as well, indicating that such power users simply let the browser running even if they do not access the Web. Unsurprisingly, the average number of page loads correlates with the average session length: the longer a session lasts the more pages a user navigates to. But note that longer sessions do not necessarily indicate that a user is browsing the Web longer than a user with a short average session length. For example, one user can browse for 1h using the 
same browser window, while another user regularly opens and closes browser windows over the same time span. To get a better picture of how idle times depend on the session lengths, Figure~\ref{fig:dobbs-median-correlation-full-vs-idle-time-explicit} shows the correlation between the average session length and the average explicit idle time for each user. Again, we use the median to represent the average values. As soon as sessions last for more than several minutes the average session length and the average idle time strongly correlates.

\begin{figure}[t!]
\centering
\subfigure[Explicit idle time]{
\includegraphics[width=0.8\textwidth]{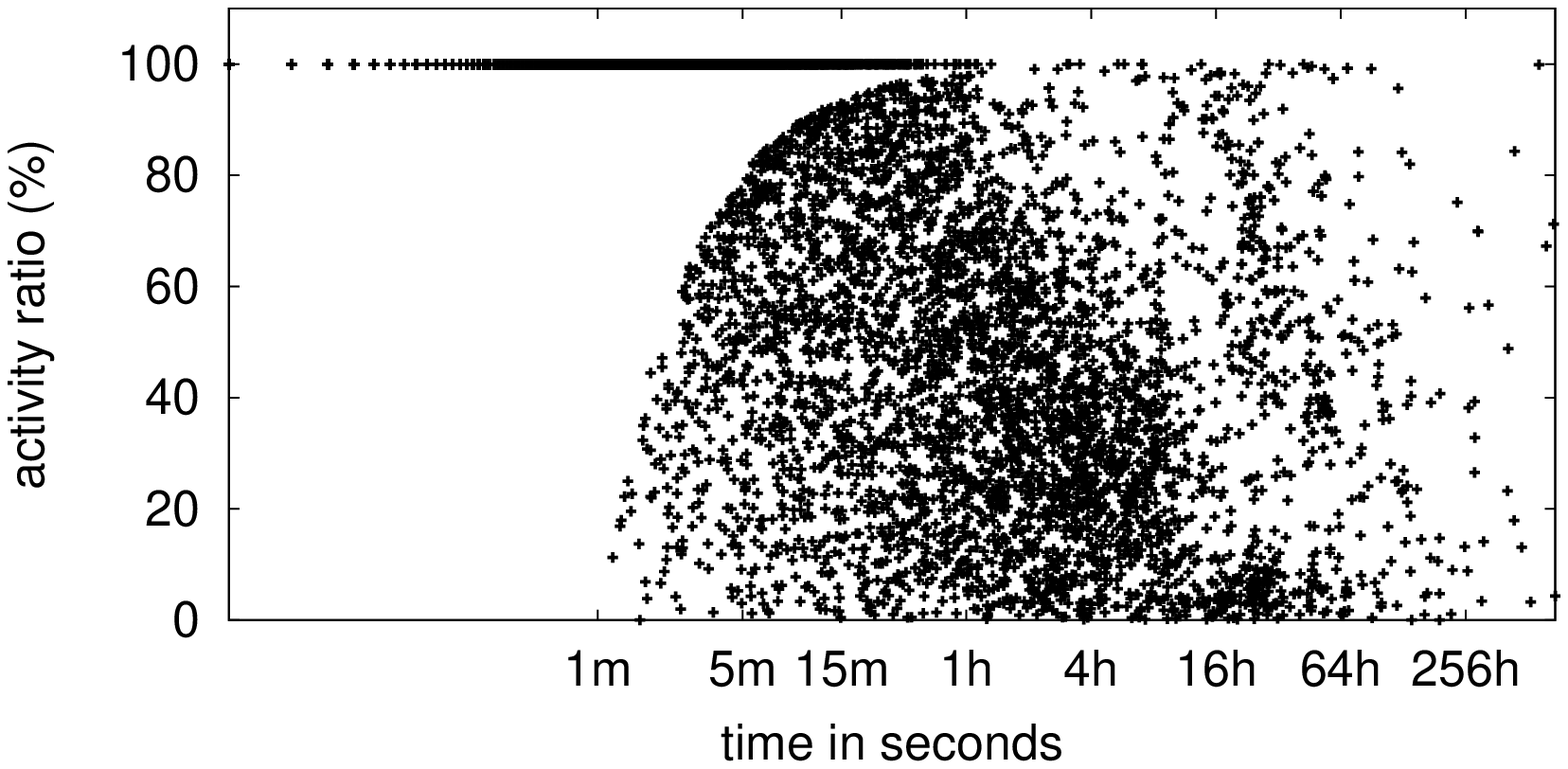}
\label{fig:dobbs-correlation-session-duration-activity-ratio-explicit}
}
\subfigure[Implicit idle time (threshold = 60s)]{
\includegraphics[width=0.8\textwidth]{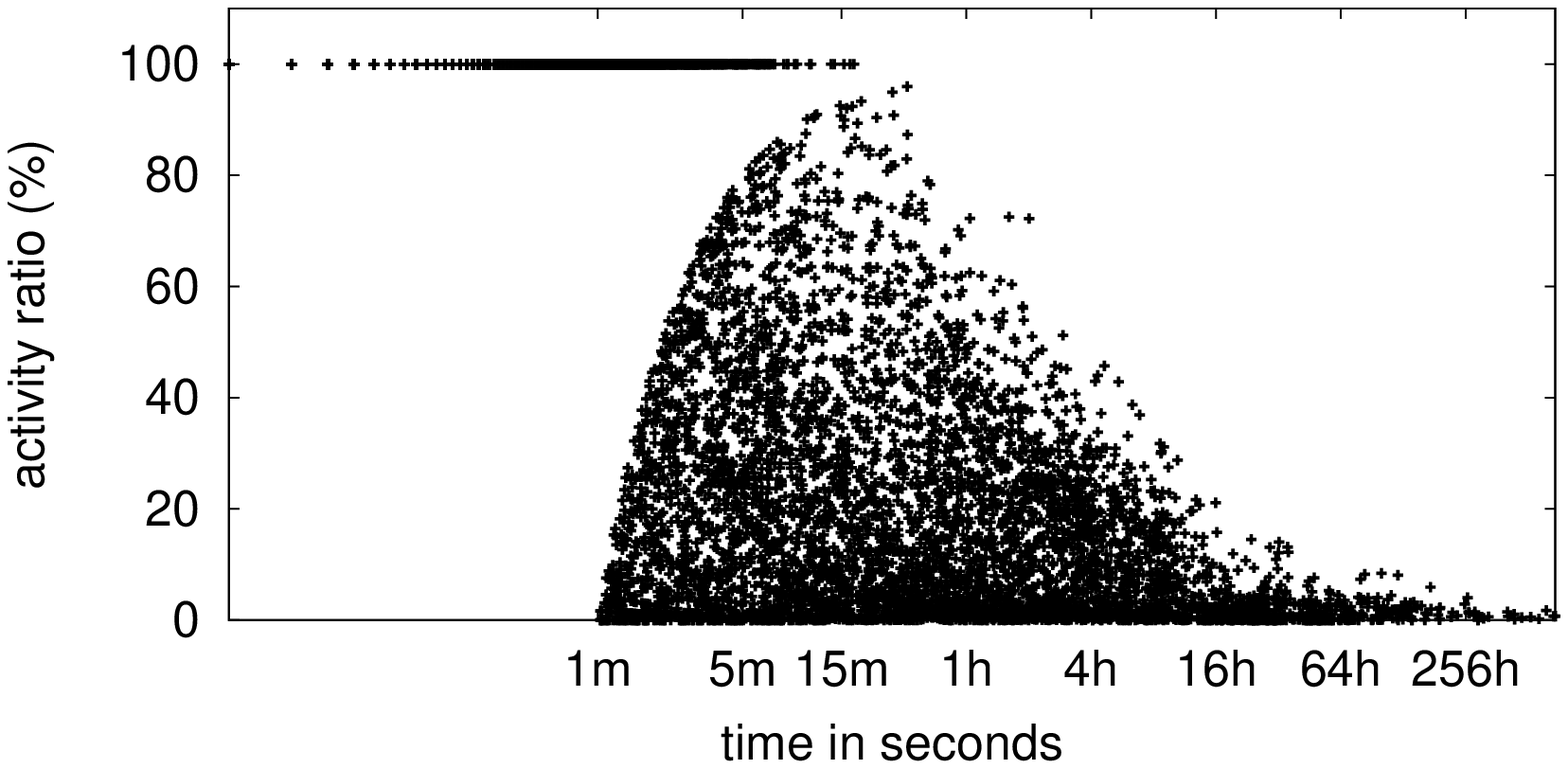}
\label{fig:dobbs-correlation-session-duration-activity-ratio-implicit-60}
}
\label{fig:dobbs-correlation-session-duration-activity-ratio}
\caption[]{Correlation between session length and activity ratio.}
\end{figure}

Figures~\ref{fig:dobbs-correlation-session-duration-activity-ratio-explicit} and~\ref{fig:dobbs-correlation-session-duration-activity-ratio-implicit-60} show the correlations between the session lengths and users' corresponding ratio of activity for each session. In general, both results show the same type of ``shark fin'' distribution (the results for implicit/240 and implicit/960 look very similar with only the start of the shark fin shifted accordingly to the right). Most naturally, during sessions shorter than the minimum idle time, a user can never be inactive. For longer session, the activity ratio of users generally decreases. As the major difference, the results for the explicit idle time show some noticeable outliers, i.e., very long sessions with an unexpected high activity ratio. We see two reasons for this. Firstly, the correct calculation of explicit idle time requires that the two required events have been logged correctly. Due to the best-effort logging, however, the dataset does contain 
corrupt data (missing events or duplicates) which might have not been handled by our applied data cleaning steps. Calculating the implicit idle time is more robust against corrupt data. Secondly, the conditions a user is considered inactive using respective browser events to calculate the explicit idle time are less strict. For example, user might read a very long article. While the user does not cause further events fetched by DOBBS, the scrolling causes the browser to consider the user as still active. Furthermore, the browser considers a user active if s/he is just moving the mouse pointer over the browser window, even if it is not the active window on the desktop. 

\begin{figure}
 \centering
 \includegraphics[width=0.8\textwidth]{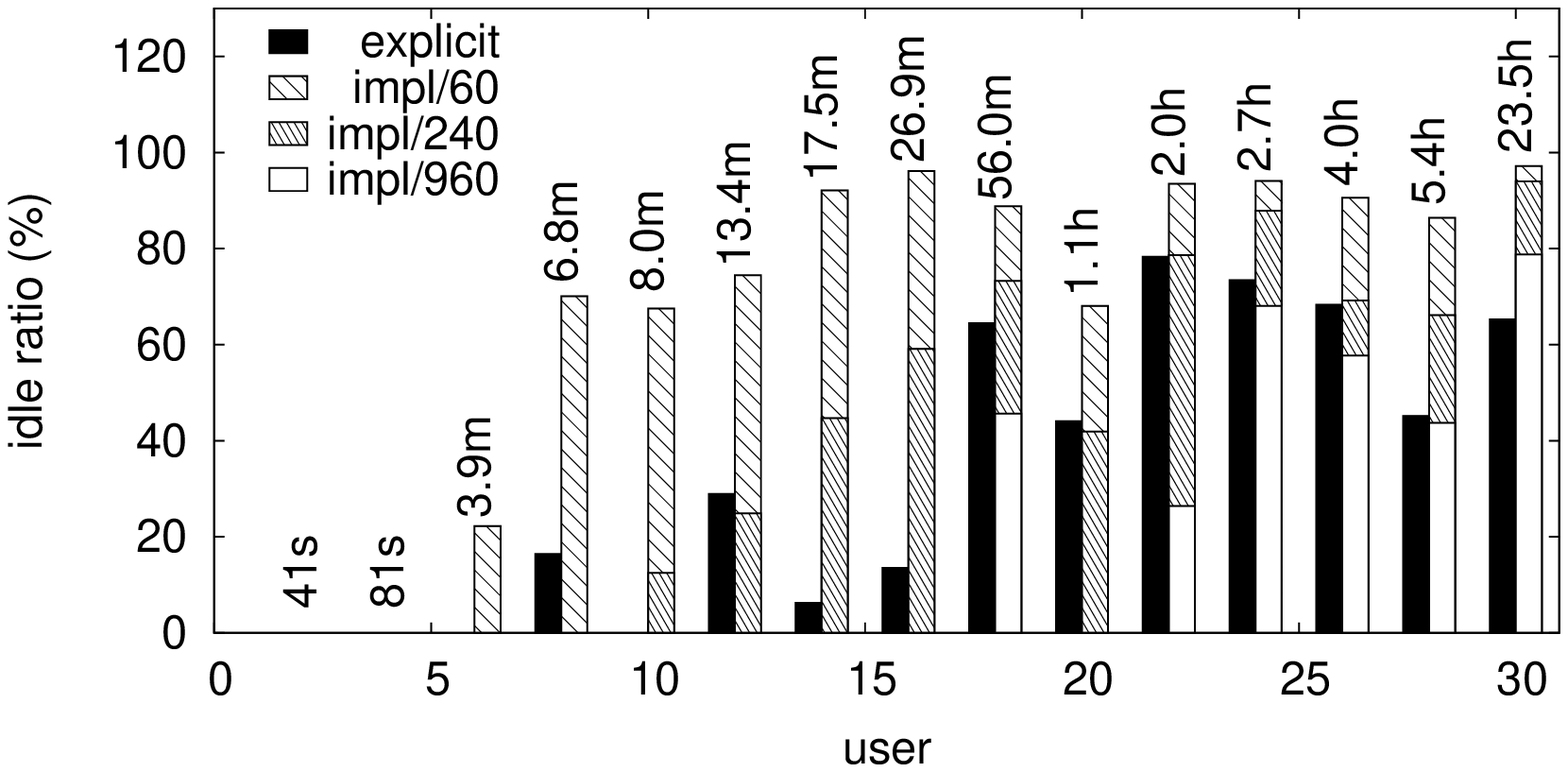}
 \caption{Comparison between explicit and different implicit idle ratios for different thresholds.}
 \label{fig:dobbs-idle-ratios-per-user}
\end{figure}

For a direct comparison between the results for the explicitly and implicitly calculated idle times, Figure~\ref{fig:dobbs-idle-ratios-per-user} shows the idle ratio based on explicit and different implicit idle time measurements. To ease presentation, we show the results only for 15 users (User 2, 4, ..., 30; ranked according to average session length). The numbers over the bars represent the average session lengths (median) for each user. In general, once sessions last longer than several minutes, the overall times of inactivity are very common and increase with increasing session lengths. This includes that the individual phases of inactivity get longer, indicated by the increasing results of the idle ratios derived from implicit idle times using larger thresholds. Particularly interesting are the results where the explicit and implicit idle rations differ significantly. This may indicate that a user is working with parallel open browser windows, or that a user is keeping a browser window open in the 
background on his/her desktop. In both case the implicit idle time increases, while the user is not considered inactive by the browser.

\begin{figure}
 \centering
 \includegraphics[width=0.8\textwidth]{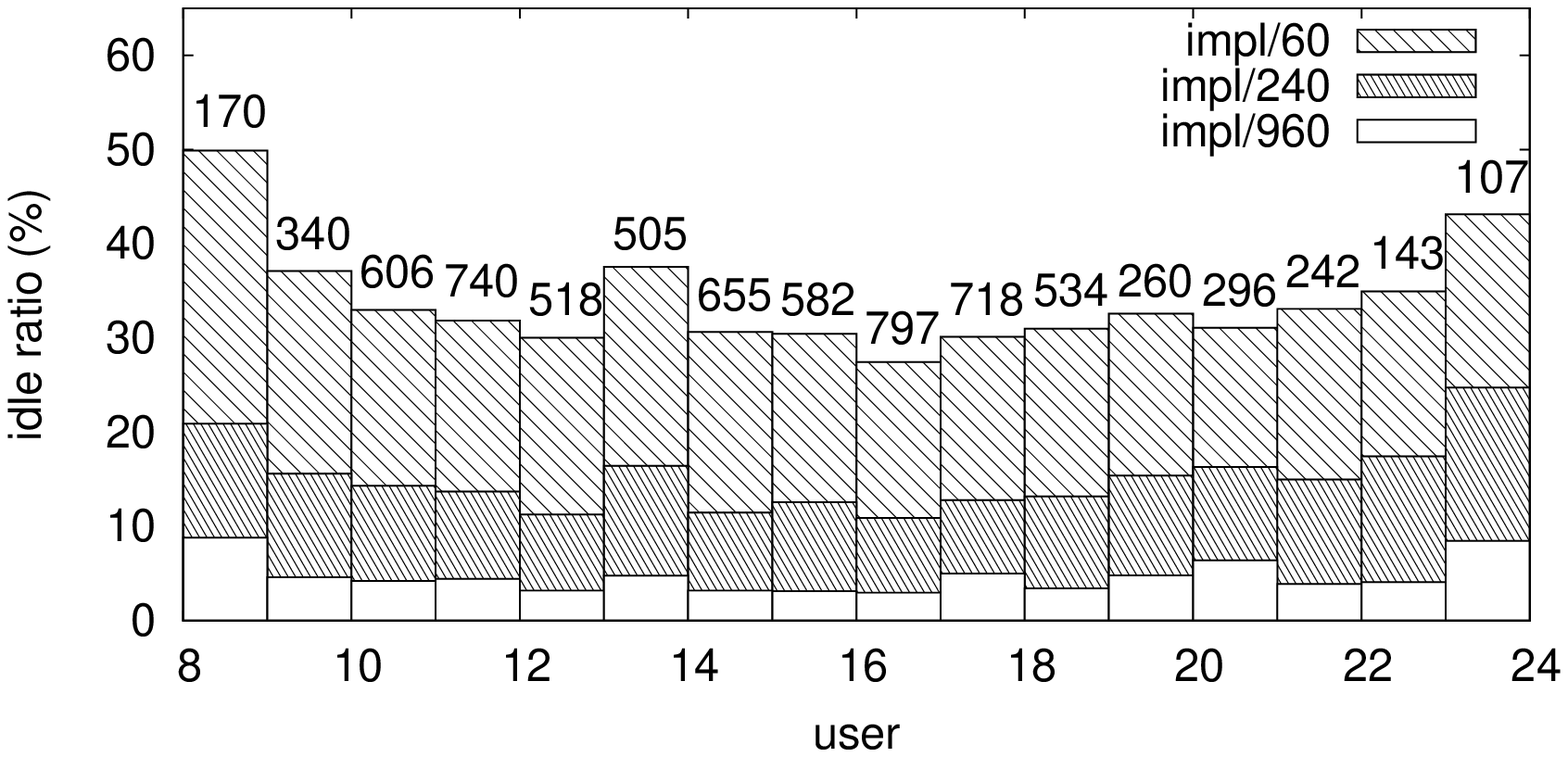}
 \caption{Distribution of idle ratios over the day for sessions shorter than 1h between 8am and 24pm.}
 \label{fig:dobbs-user-activity-over-day-implicit-all}
\end{figure}

Lastly, we looked at how the idle ratio changes over the day. For this, we grouped all sessions shorter than 1h according to their start time and calculated the average idle ratio for each hour of the day. Figure~\ref{fig:dobbs-user-activity-over-day-implicit-all} shows the results for 8am until 24pm for different implicit idle times; the numbers over the bars represent the number of sessions within that group. We omitted the interval from 0-7am since these groups contained less than 100 sessions. Since we assume that most users access the Web as part of their work -- see Section~\ref{sec:discussion} for details -- all figures reflect the typical working intervals: from 9-12am and 2-5pm the number of sessions per hour are the highest and the average idle ratios are the lowest. Even the common time for a lunch break (1-2pm) is clearly visible.

Summing up, all results consistently shows that idling during an on-going browsing session is very common. On the other hand, there are various means to calculate the time a user is inactive, with different approaches reflecting different aspects of ``being inactive'' and typically yielding different results. Which measure or parameter to use for quantifying the activity of users depends on the specific research questions behind an analysis.

\subsection{Website Popularity}
As previous results show, both parallel and passive browsing are common phenomenons. We now investigate how these two aspects of online browsing behavior potentially allow for novel types of metrics to quantify the popularity of websites. We first looked at the effects of parallel and passive browsing on the 100 most visited websites. For this, we summed up the aggregated the loaded and focused times of pages of the same website, i.e., pages with the same domain. In the following results, we also highlighted a set of selected websites to give some examples.

\begin{figure}[t]
 \centering
 \includegraphics[width=.8\textwidth]{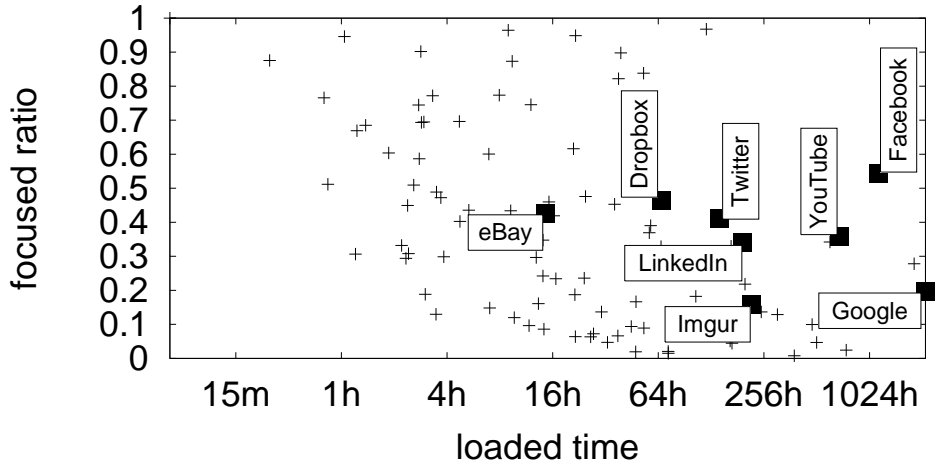}
 \caption{Correlation between loaded and focused times for popular domains.}
 \label{fig:dobbs-loaded-time-vs-focused-ratio-all}
\end{figure} 

\begin{figure}[t]
 \centering
 \includegraphics[width=.8\textwidth]{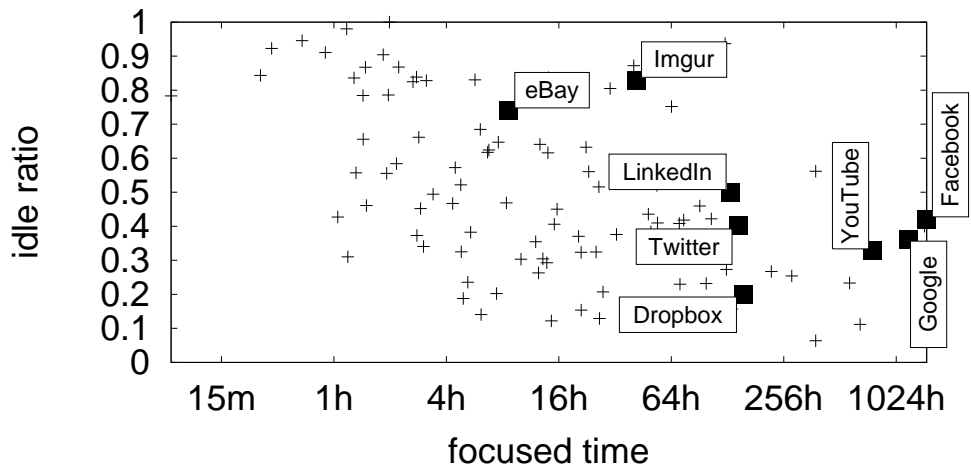}
 \caption{Correlation between and focused times and idle-corrected focused time.}
 \label{fig:dobbs-focused-time-vs-correct-focused-time-ratio-all}
\end{figure}

Figure~\ref{fig:dobbs-loaded-time-vs-focused-ratio-all} plots the time a website was loaded in a browser window against the focused ratio, i.e., the relative time the website was visible in the selected tab. Note that here we consider only the times when users were considered active. The results indicate that the focused ratio can differ significantly between different websites, independently from their popularity in terms of overall loaded time. For example, \textsc{Google} shows a rather low focused ratio. Our explanation is that many users open and look at result pages while keeping the \textsc{Google} result list in a background tab. \textsc{Facebook} pages, on the other hand, are in relation much longer on display. Figure~\ref{fig:dobbs-focused-time-vs-correct-focused-time-ratio-all} compares the overall time the pages of website were focused (i.e., visible in the selected tab) and the overall time the pages were focused \textit{and} the users were active during the time.  Again, the ratios differ 
significantly between websites. Here, \textsc{Google} and \textsc{Facebook} yield similar result, i.e., both sites are equally left on display while user are idling. In contrast, user browse the image sharing site \textsc{Imgur} with far fewer phases of inactivity.

\begin{table*}[htb]
 \centering
 \small
 \begin{tabular}{|c|l|l|l|l|l|l|}
  \hline 
  \textbf{}	& \textbf{Alexa}		& \textbf{visit time ratio}	& \textbf{page load ratio}	& \textbf{focused ratio}	& \textbf{active ratio} \\
  \hline\hline
  1		& \textsc{Google} (1)		& \textsc{Google} (10.5\%)	& \textsc{Facebook} (14.3\%)	& \textsc{Facebook} (54.5\%)	& \textsc{LinkedIn} (50.0\%) \\
  \hline
  2		& \textsc{Facebook} (2)		& \textsc{Facebook} (6.4\%)	& \textsc{Google} (10.7\%)	& \textsc{Twitter} (41.0\%)	& \textsc{Facebook} (42.1\%) \\
  \hline
  3		& \textsc{YouTube} (3)		& \textsc{YouTube} (4.3\%)	& \textsc{YouTube} (5.5\%)	& \textsc{YouTube} (35.8)	& \textsc{Twitter} (40.2\%) \\
  \hline
  4		& \textsc{LinkedIn} (11)	& \textsc{LinkedIn} (1.2\%)	& \textsc{LinkedIn} (2.2\%)	& \textsc{LinkedIn} (34.0\%)	& \textsc{Google} (36.2\%) \\
  \hline
  5 		& \textsc{Twitter} (12)		& \textsc{Twitter} (0.8\%)	& \textsc{Twitter} (1.4\%)	& \textsc{Google} (19.7\%)	& \textsc{YouTube} (32.8\%) \\
  \hline
 \end{tabular}
 \caption{Ranking of popular websites according to different metrics.}
 \label{tab:ranking-metrics}
\end{table*}
Given thesere sults, we lastly investigated how the time a page was actually visible and the time users were inactive can be used to derive novel metrics to quantify the popularity of a website. For this, we first selected the 5 websites that were commonly visited by most of the DOBBS users: \textsc{Facebook}, \textsc{Google}, \textsc{LinkedIn}, \textsc{Twitter}, and \textsc{YouTube}. While many other sites have also been visited very frequently, this has been done only by a small set of users for each site. We then ranked these sites according to different metrics; Table~\ref{tab:ranking-metrics} show the results. In the following, we explain the various metrics in more detail and discuss briefly the rankings the metrics yield.

\textit{(a) Alexa rank:}
As baseline ranking, we used the list of top sites according to Alexa.\footnote{http://www.alexa.com/topsites (February 10th, 2014)} While the Alexa ranking is said to be biased -- for example, the required toolbar is installed and promoted mostly by webmasters to pitch their rank -- we are less interest in the resulting ranking. Here, we only want to highlight how the ranking changes with respect to different metrics. All of the 5 selected websites are within the top 12 sites according to Alexa, meaning that we do not have included any ``odd'' site in our selection.

\textit{(b) Loaded time ratio and hit ratio:} The loaded time ratio represents the distribution of times the users have visited pages of a website. The higher the value the longer users had a website open in their browser, in either the selected or a background tab. Similarly, the ratio of page hits refers to the number of page hits as classic website popularity metric. Again, we use the ratio of hits, i.e., the number of pages of a website a user requested divided by number of all page requests by a user, to calculate the average value across all users. Both metrics yield very similar rankings compared to baseline ranking. After all, among others, Alexa uses the number of user visits as metric which essentially are the page hits and tend to correlate with the time a page has been loaded.

\textit{(a) Page focused ratio:} The focused ratio reflects the difference between the times the pages of a domain have been loaded in a browser tab and the time those pages were actually visible in the currently selected tab. We argue that this ratio kind of quantifies how ``absorbing'' a website is, and that it might be of particularly interest for advertisers to ensure that their ads are indeed on display. In this context, given two sites with otherwise similar characteristics (page hits, unique visitors, etc.), the one with a higher focused ratio is more attractive for placing ads. As mentioned above, that \textsc{Google} performs rather worse according to this metric is not unexpected, since users often keep \textsc{Google} result lists in a background tab. On the other hand, users give a rather high attention to social network platforms such as \textsc{Facebook} and \textsc{Twitter}.

\textit{(d) Active ratio:} The active ratio compares the time the pages of a domain had the focus (i.e., the pages were visible in the selected tab) and the time the pages had the focus \textit{and} the user was actually considered active. We consider this metric to be useful for quantifying how ``engaging'' a website is. This can be of major interest for webmasters who aim to increase visitor or member loyalty. Again, social network platforms elicit the higher activity from users in terms of, for example, browsing profiles, as well as reading, writing or rating posts. Not unintuitively, \textsc{YouTube} shows the lowest value. We presume that it is pretty common that users sit back and watch a video clip without any browser interactions during the clip's runtime. Even more, users might just listen to music clips and leave their computer altogether. As such, the activity ratio is only a good metric, if the level of engagement is indeed an important aspect for a website. For example, we assume that online radio platforms would yield even lower values (the current dataset does not include enough page views to online radio sites made by enough users to allow for a meaningful analysis).

Summing up, the rankings induced by the \textit{focused ratio} and \textit{active ratio} differ clearly from more traditional approaches. We therefore argue that both measures do provide good reasons to re-think the methods for quantifying the popularity of websites. We also want to emphasize, however, that considering these metrics has to be done with great care. The interpretation of the results about how ``absorbing'' or ``engaging'' a website depends on the purpose of the site. For example, an online radio site running for hours on a background tab will result in very low values for both metrics.

\section{Discussion}
\label{sec:discussion}
To the best of our knowledge, DOBBS offers a rather unique dataset for investigating the online browsing behavior of Web users. As such, we believe that our results provide novel and useful insights. While the dataset allows for very broad spectrum regarding users' browsing behavior, we focused on the concepts of parallel and passive browsing, and their potential effects on quantifying the popularity of websites. In this section, we discuss our evaluation results with respect to generalizability and their impact on different Web research areas and Web development.
\\
\\
\textit{Limited sample size.}
One of the biggest challenges for DOBBS is to motivate user to contribute by installing the add-on. Firstly, the add-on does not provide an added value to users, and despite the anonymisation and application of encryption techniques, users might perceive privacy risks. At the time of writing, 139 users have installed the add-on for Firefox or Chrome. However, not all users show any frequent or recent activity, indicating that they removed the add-on again. The 30 selected users for our evaluation show a constant browsing activity over time. While an evaluation would benefit from a larger user base, we argue that 30 is a reasonable sample size for a first statistical analysis of the dataset.
\\
\\
\textit{Limited demographic.}
Although we cannot know for sure, due to the anonymisation of the data, we assume that most DOBBS users have an academic background. The reason for that assumption is our current choice of channels to announce and promote the project. This mainly includes emails to popular mailing lists in the areas of databases, information systems, data analytics, and human-computer interactions. We also handed out flyers and made announcement during talks at conferences. As such, we assume a rather tech-savvy set of users who use the Web on a daily basis for personal as well as professional reasons. Hence, we do not claim that our result can be generalized to a more average demography of Web users. For example, from a typical member of the ``Facebook Generation'' we would expect less visits on programming Q\&A sites such as \textsc{StackOverflow}, but maybe more visits on online radio or online games platforms.
\\
\\
\textit{Interpretation of results.}
DOBBS is a designed as a field study to elicit normal browsing behavior of users, with no interference from any controlling entity. Furthermore, the best-effort logging of the add-on may cause incomplete logs or even duplicate entries in the dataset. If an analysis of the dataset does not (heavily) depend on these types of corrupt logging data they can be ignored. As preliminary step before our evaluation, we therefore applied multiple data cleaning steps to extrapolate missing data and to remove duplicate entries. The granularity of the DOBBS dataset makes these steps valid and applicable for most evaluations. Still, every so often, some results can be interpreted as outliers, for example, when a user has viewed individual pages for several hours at a time (even excluding possible phases of inactivity). While these rather unexpected results can indeed be valid, we are more inclined to consider them as artifacts stemming from corrupt data. Note that often only a single corrupt event entry can result 
in such an outlier. For a statistical analysis, these cases typically pose no problems. Firstly, after some basic data cleaning steps, really peculiar results are very rare. And secondly, since the distributions of analyzed parameters are typically skewed, we use the median to average the result, which is robust against rare outliers.
\\
\\
\textit{Impact of our results and on-going work.}
Having insights into the browsing behavior of users allow scientists as well as software engineers from a large variety of disciplines to derive both fundamental and applied knowledge from different perspectives: 

\textit{(1) Redefining the popularity of websites.} That the number of page views is not necessarily a perfect metric to quantify the popularity of websites has long been acknowledged, and a wider range of alternative metrics have been introduced (e.g., number of unique visitors, repeated visits, bounce rate, etc.). More recently, particularly the dwell time, i.e., the time a user stayed on a page or site, gained a lot of interest. The information how long and how active users visit a website provides implicit feedback about its quality, potentially improving ranking algorithms of Web search engines. For example, a site that typically resides in a background tab might be considered differently than sites that typically involve more active user participation. However, as our results show, using the dwell time as popularity metric is challenging since calculating this time is not trivial. Server-side solutions can only estimate how long a user stayed on a page. But also client-side approaches have to address various pathological cases that might negatively affect the results. For example, a user might ``forget'' a page in a background tab. Also, the effect of being in a background tab depends on the type of website -- an online radio site does not need to have the focus for the user to enjoy the content.

\textit{(2) Improved Web browser design.} With the browser being the main application to access the Web, knowledge about browsing behavior enables the design and development of new features that improve the online experience of users. Examples include the hiding of and the quick access to tabs containing passively used webpages (e.g., online radio), or the automatic re-arranging of tabs according to their usage. Furthermore, new optimization approaches to save bandwidth or computing resources are conceivable. This might include, for example, special ``idle modes'' for browsers or individual tabs where dynamic pages are not automatically updated if, e.g., the tab is in the background or the user is recognized as inactive.

\textit{(3) The Web and society.} Understanding how and when users browse the Web also allows deriving statements on why they use it (e.g., for information seeking, entertainment, socializing, or for other work or leisure-related activities). This in turn provides elusive insights into the sociological impact of the Web -- that is, how the Web and ``being online'' shapes peoples' life. For example, the DOBBS dataset already contains some what might be called power users with browsing session of sometimes several days (note that this naturally implies that they do not switch of their computers for that long, if at all). This indicates that the so-called ``Always On'' lifestyle does more or more penetrate modern societies.

Our evaluation yields several novel results regarding users' browsing behavior. In the scope of this paper, we have limited ourselves to rather high-level results. The range and granularity of the logged data allows for a broader spectrum and more detailed analysis. For example, the dataset contains the information when a browser window was in the background and/or even minimized. This in turn might affect the results regarding the time a user has actually viewed a page. For our evaluation, we argue -- and looking at the data justifies this -- that the window is minimized and/or in the background correlates with the time a user is inactive. Furthermore, with respect to parallel browsing, the DOBBS dataset enables to retrace a users' navigation between pages. e.g., how a user switched between different open tabs or for how many page loads s/he actually uses a tab. In general, the aspects of the dataset and the level of detail that drive an evaluation depends on the specific research questions.

\section{Conclusions}
\label{sec:conclusions}
In this paper, we presented the results of a first analysis of the DOBBS dataset with logging data stemming from a sufficient number of users. Compared to similar available efforts, DOBBS tracks users' browsing behavior with a high granularity while being designed as a long-term field study to elicit true every-day browsing behavior. Due to the basic design choice of employing a best-effort logging, the dataset features incomplete as well as corrupt data. We observed, however, that these pathological cases are rare. Due to the detailed data structure of the logged events it is easy to identify and filter out corrupt data. Also, if needed, one can estimate missing data in a meaningful manner. The collected data cover a wide range of aspects regarding the browsing behavior of Web users. For this evaluation, we focused on parallel and passive browsing. We found that both phenomenons are very common but differ significantly between users. Significant differences also showed up when we compared the level 
of parallel and passive browsing between different websites. Based on this observation, we investigated how such knowledge may provide novel approaches to quantify the popularity of websites. More specifically, we introduced the \textit{focused ratio} and \textit{activity ratio} as new metrics towards the notions of how ``absorbing'' or how ``engaging'' a website is. For a set of widely known websites, we could show that both metrics do indeed induce very different rankings compared more traditional methods. 

\begin{figure}[t!]
 \centering
 \includegraphics[width=0.8\textwidth]{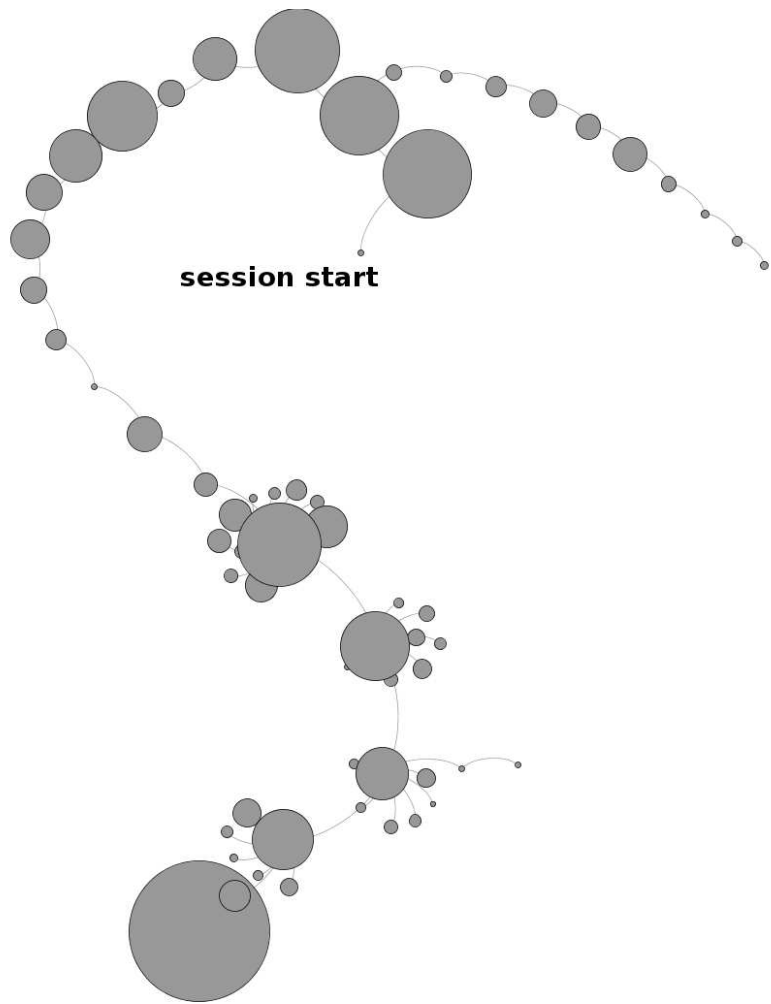}
 \caption{Example graph representation of a browsing session. The size of a node reflects the duration the user spent on the corresponding web page.}
 \label{fig:gephi-414431039}
\end{figure}

While the results of our analysis already provide interesting insights, they represent just our first step towards a more complete picture of the browsing behavior of online users. Not only does the DOBBS dataset allows for the investigation of a much wider range of aspects regarding browsing behavior, also the phenomenons of both parallel and passive browsing can be studied in more detailed. The dataset features sufficient logging data to accurately re-trace a user's navigation between pages using tabbed browsing. To give a brief outlook, the navigation paths using multiple tabs can then be represented as a graph; Figure~\ref{fig:gephi-414431039} shows an example. In doing so, the browsing behavior can be depicted as a directed tree with the root being the startup of the browser. To meaningfully analyze these trees to, e.g., categorize different types of parallel browsing, requires appropriate measures, including more sophisticated one from graph theory (e.g., the branching 
factor, the average shortest path from the root). Regarding passive browsing, one can further refine the calculation of the idle time of users by considering if a browser window was in the desktop background or even minimized. Answering those and closely related research questions represent our immediate on-going work.

\newpage
\renewcommand{\baselinestretch}{0.97}
\bibliographystyle{abbrv}
\bibliography{literature}

\end{document}